\DocumentMetadata{}
\documentclass[sigconf,authorversion,nonacm]{acmart}

\usepackage{cleveref}
\usepackage{subcaption}
\usepackage{natbib}

\newcommand*\circled[1]{{\textbf{(#1)}}}

\AtBeginDocument{%
  \providecommand\BibTeX{{%
    \normalfont B\kern-0.5em{\scshape i\kern-0.25em b}\kern-0.8em\TeX}}}

\begin{document}

\title{How Deep Is Your Gaze? Leveraging Distance in Image-Based Gaze Analysis}


\author{Maurice Koch}
\email{maurice.koch@visus.uni-stuttgart.de}
\orcid{0000-0003-0469-8971}
\affiliation{%
  \institution{University of Stuttgart}
  \city{Stuttgart}
  \country{Germany}
}

\author{Nelusa Pathmanathan }
\orcid{0000-0002-6848-8554}
\email{nelusa.pathmanathan@visus.uni-stuttgart.de}
\affiliation{%
  \institution{University of Stuttgart}
  \city{Stuttgart}
  \country{Germany}
}

\author{Daniel Weiskopf}
\orcid{0000-0003-1174-1026}
\email{daniel.weiskopf@visus.uni-stuttgart.de}
\affiliation{%
  \institution{University of Stuttgart}
  \city{Stuttgart}
  \country{Germany}
}

\author{Kuno Kurzhals}
\orcid{0000-0003-4919-4582}
\email{kuno.kurzhals@visus.uni-stuttgart.de}
\affiliation{%
 \institution{University of Stuttgart}
 \city{Stuttgart}
 \country{Germany}
}

\renewcommand{\shortauthors}{Koch et al.}

\begin{abstract}
  Image thumbnails are a valuable data source for fixation filtering, scanpath classification, and visualization of eye tracking data.
  They are typically extracted with a constant size, approximating the foveated area.
  As a consequence, the focused area of interest in the scene becomes less prominent in the thumbnail with increasing distance, affecting image-based analysis techniques.
  In this work, we propose depth-adaptive thumbnails, a method for varying image size according to the eye-to-object distance.
  Adjusting the visual angle relative to the distance leads to a zoom effect on the focused area.
  We evaluate our approach on recordings in augmented reality, investigating the similarity of thumbnails and scanpaths. 
  Our quantitative findings suggest that considering the eye-to-object distance improves the quality of data analysis and visualization.
  We demonstrate the utility of depth-adaptive thumbnails for applications in scanpath comparison and visualization.
\end{abstract}




\maketitle

\section{Introduction} \label{sec:intro}
Video-based eye tracking methods for wearable devices often include a video of the world-view perspective to interpret viewing behavior and identify potential areas of interest (AOIs) in an environment. Since this video is often recorded by a single camera, gaze data is mapped after calibration to the 2D image plane of the video, resulting in the loss of depth information of the environment. Hence, the recorded data of a complex 3D environment is often reduced to a time series of gaze samples with $x$- and $y$-coordinates relative to the calibrated reference system. 

Based on this data, numerous techniques have been developed over the years to analyze gaze and stimulus data using statistics~\cite{Holmqvist2011} and visualization~\cite{Blascheck2017}: AOIs are defined and mapped to fixations, either by hit detections on polygonal shapes drawn into the video~\cite{kurzhals2014}, or by image-based approaches~\cite{kurzhals2016visual,pontillo2010} that extract thumbnails of the area around a gaze point.
Drawing polygons has the advantage that they are directly visible in the stimulus and a video playback shows how they change in position and size. However, annotating and tracking changing polygons is time-consuming and might vary in the accuracy of the defined shapes.
Image-based approaches work directly with the stimulus, AOIs solely serve as labels that are assigned to a thumbnail. This approach can reduce annotation effort and makes it easier to compare multiple annotators to calculate their agreement. Further, the extracted thumbnails can be used directly for training AOI classifiers~\cite{Castner2020}, compare scanpaths~\cite{Koch2022}, and visualize image similarities to identify and annotate AOIs efficiently~\cite{Kurzhals2021}. 

One issue with this technique is that thumbnails only show parts of the stimulus, making it difficult to identify specific AOIs without knowing the context. Additionally, since thumbnails are always taken with a constant size from the image plane, an object will occur in different sizes, depending on the distance to the camera.
Although this is natural due to perspective foreshortening, such differences impair image-based processing and might result in false~classifications.

In this work, we address this problem by including depth information to determine the size of the thumbnail, mainly to keep the visible part of an object consistent to improve image-based processing techniques for eye tracking recordings. Technically, we apply a zoom on the image, resulting in a decreased crop area for objects further away. We investigate how this modification of the thumbnail extraction process influences techniques such as scanpath comparison and the visualization of the data.

More specifically, our contributions are:

\begin{itemize}
    \item A new method for depth-adaptive image thumbnails based on eye-to-object distance, which we examine in augmented reality (AR) environments.
    In general, our method is hard\-ware-agonistic and similarly applies to non-AR devices such as mobile eye tracking glasses.
    The only formal requirement is a continuous measurement of eye-to-object distances, which can, for instance, be obtained by stereo camera matching or RGB-D sensors.
    \item We evaluate our approach in two scenarios common in eye tracking analysis: (1) scanpath similarity measurements and (2) gaze visualization. 
\end{itemize}
Overall, our results show that depth-adaptive thumbnails provide a more robust basis for image-based analysis of gaze data than the fixed-size baseline approach.
In the future, this technique could provide better performance for live classification and post-experimental analysis of eye movements. 

\section{Related Work} \label{sec:related}

In contrast to techniques that solely rely on properties of the recorded gaze data, such as spatial position and velocity of eye movements, image-based techniques include valuable information about the visual stimulus to improve analysis.
From a data processing perspective, the extracted thumbnails can be applied to various kinds of analysis.
For instance, \citet{Kurzhals2016} represented the scanpaths of participants by image sequences and used their similarity as input for established scanpath comparison metrics, showing that groups of similar behavior can be identified without the necessity to annotate AOIs in advance. 
A similar approach was used by \citet{Castner2020} to classify scanpaths of novice and expert dentists.
\citet{Steil2018} computed the visual similarity of image patches from eye tracking videos to detect fixations.

Different visualizations have been presented to facilitate the interpretation of these techniques. 
\citet{Kurzhals2016} showed all gaze samples of a participant as a long sequence of thumbnails. 
Later, the authors adapted this idea for fixations, reducing the visual load on the user~\cite{kurzhals2016fixation}. 
The authors also used thumbnails for visualization techniques to support AOI annotation~\cite{kurzhals2016visual,Kurzhals2021}. 
As discussed, such approaches are based on a fixed visual angle, often approximating the foveated area with some additional offset. 

We argue that this static representation leads to issues with the previously discussed techniques, which we address with our proposed depth-adaptive thumbnails.
We will further elaborate on this assumption in Section~\ref{sec:eval}.

A multitude of works investigated how depth information can improve the user experience of AR and virtual reality (VR) applications. Applications include resolving target ambiguity \cite{Mardanbegi2019}, controlling see-through vision \cite{WangZhimin2022}, rendering on-demand content \cite{Arefin2022}, and attention detection on small targets \cite{Sidenmark2023}. Previous work has also investigated attentional maps and gaze plots in 3D \cite{Pfeiffer2016, Stellmach2010} but to the best of our knowledge, no work so far explicitly leveraged depth information for visualization purposes. In this paper, we explore opportunities for depth estimation on image thumbnails.

\section{Technique} \label{sec:technique}

\begin{figure*}
    \centering
    \includegraphics[width=1\textwidth]{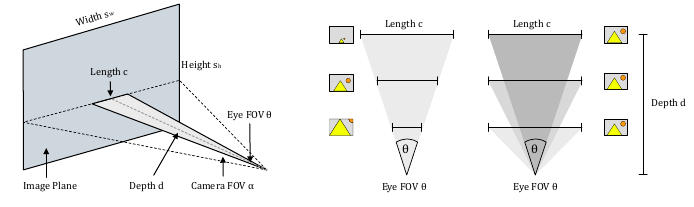}
    \caption{A fixed field of vision, given by angle $\theta$, causes scene objects to become cropped when they are too close to the eye or to vanish at great distances. Setting the actual length $c$ to a constant value keeps the scene objects in the focus of the beholder by varying the eye's field of vision.}
    \Description{An illustration of the proposed depth-adaptive thumbnails composed of two parts. On the left, relevant geometric units relevant to this paper are introduced. It depicts the geometric relationship between the dimension of the image plane, visual angle, gaze depth, camera field-of-view, and the actual length. On the right, the basic concept of depth-adaptive thumbnails is illustrated using the previously introduced terminology.}
    \label{fig:concept1}
\end{figure*}

Before we present our approach, we revisit the geometric relationships between visual angle and actual length as depicted in \Cref{fig:concept1}.
Given an eye's visual angle $\theta$, the actual length $c$ is proportional to the eye-to-object distance $d$ 
\cite{Holmqvist2011}:

\begin{equation} \label{eq:c}
    c = 2 \cdot \tan \left (\frac{\theta}{2} \right ) \cdot d
\end{equation}
A reoccurring task is to compute the actual length of a screen in stationary eye tracking experiments. For example, 
assuming foveal vision with $\theta = 2^\circ$ and a distance of $d = 0.5$\,m to the screen, the actual length is $c \approx 1.7$\,cm.

\paragraph{Classic Patch}

To derive the gaze patch size, we need to convert length units (meters) to screen coordinates (pixels).
To this end, we additionally need the dimension $s_w \times s_h$ (in meters) and resolution $r_x \times r_y$ (in pixels) of the screen.
In mobile eye tracking, the reference coordinate system is typically given by the front-facing camera, so $s_w \times s_h$ refers to the size of the image plane.
Computing this size requires the front-facing camera's horizontal $H$ and vertical field of view $V$.
Then, $s_w$ and $s_h$ are calculated using \Cref{eq:c} by setting $\theta = H$ and $\theta = V$, respectively.
The actual size in pixel units is then given by:

\begin{equation} \label{eq:cp2}
    \begin{split}
        C_x(\theta) &= \frac{\tan \left (\frac{\theta}{2} \right )}{\tan \left (\frac{H}{2} \right )} \cdot r_x \\
        C_y(\theta) &= \frac{\tan \left (\frac{\theta}{2} \right )}{\tan \left (\frac{V}{2} \right )} \cdot r_y
    \end{split}
\end{equation}
The gaze patch is then given by cropping a rectangular region $C_x(\theta) \times C_y(\theta)$ from the image plane around the gaze point of interest.
It is important to notice that the depth $d$ cancels out in \Cref{eq:cp2}, which means in pixel space the actual size is independent of the depth.

\paragraph{Depth-adaptive Patch}

Using \Cref{eq:c} to determine the patch size leads to two potential problems: (1) far-off objects may vanish and as a result become unrecognizable in the image plane, and (2) very close objects may be clipped because they extend outside of the image plane.
We argue that image-based analysis methods are potentially impaired by both of the aforementioned problems.
Instead, we vary the visual angle $\theta$ based on the eye-to-object distance to keep the actual length $c$ constant.
Similarly to \Cref{eq:cp2}, the resulting patch size in units of pixels is given by:
\begin{equation} \label{eq:cp3}
    \begin{split}
        D_x(c) &= \frac{c}{2d \cdot \tan \left (\frac{H}{2} \right )} \cdot r_x \\
        D_y(c) &= \frac{c}{2d \cdot \tan \left (\frac{V}{2} \right )} \cdot r_y
    \end{split}
\end{equation}

We implemented both \Cref{eq:cp2} and \Cref{eq:cp3} in Python~3.9 \shortcite{python} with OpenCV \shortcite{opencv} binding to perform the image processing.
The results in \Cref{sec:eval} were produced by a combination of PyTorch \shortcite{pytorch}, OpenCV, and open-source implementations of the referenced methods.

\section{Evaluation} \label{sec:eval}

\begin{figure*}
    \centering
    \includegraphics[width=1\textwidth]{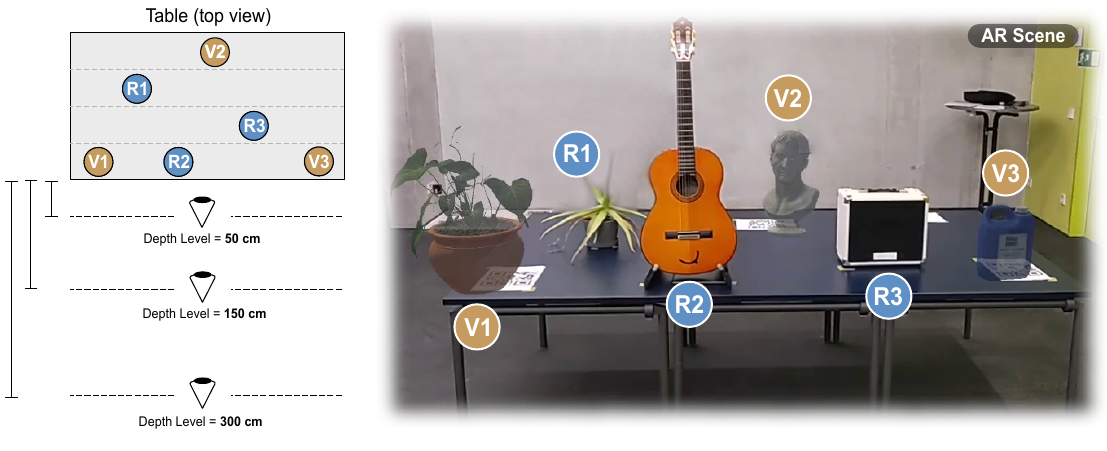}
    \caption{Benchmark scene comprised of three real-world objects (R1 --- R3) and three virtual objects (V1 --- V3) on a desk. Left:~Experiment space with three viewing locations positioned at distances of 50\,cm, 150\,cm, and 300\,cm.  Right: Scene view from the HoloLens 2.}
    \Description{The right image shows a first-person view of the AR setup. The scene is comprised of three virtual (potted plant, head of a sculpture, canister) and three real-world objects (Aloe Vera plant, guitar placed on a stand, guitar amplifier) placed on a table. The left illustrates a top-down view of the same AR setup. Additionally, distance markers indicate from which positions the participants performed the tasks.}
    \label{fig:setup}
\end{figure*}

We evaluate our approach for two typical analysis scenarios in eye tracking data analysis: scanpath similarity and scanpath visualization.
In the following, we specify the methods to be evaluated and outline the evaluation procedure for both scenarios. 

\paragraph{Scanpath Similarity} We evaluate our approach on the algorithm by \citet{Smith1981} applied to sequence of image patches similarly to \citet{Castner2020}.
We employ the cosine similarity on 512-D feature vectors extracted from ResNet18 \cite{He2016} to compute the substitution costs in the Levenshtein distance~\cite{Levenshtein1966} and the Smith-Waterman algorithm.
We choose a gap penalty of $0.5$ in the Smith-Waterman algorithm.
For results on the Levenshtein distance, we refer to our supplemental material.

\paragraph{Scanpath Visualization} Image patches are the basis of Gaze Stripes \cite{Kurzhals2016}, an image-based visualization of scanpaths.  Image patches also have been used for interactive annotation of AOIs in mobile eye tracking recordings. For example, \citet{Kurzhals2021} generated a 2D gaze scatter plot using dimensionality reduction (DR) on image patches. We evaluate the impact of our approach on this visualization both quantitatively and qualitatively.

\subsection{Benchmark Dataset}

We equipped four participants with a Hololens 2 to look at real and virtual objects from three different distances.
For gaze and video recording, we used the Augmented Reality Eye Tracking Toolkit for Head Mounted Displays (ARETT) \cite{Kapp2021}.
We obtained fixations using dispersion-based fixation classification provided by ARETT's \texttt{R} package.
Our study received approval from the ethics committee of our institution.
We have made this benchmark dataset openly accessible \cite{darus-4141_2024}.
In the following, we outline task and scene constraints to study the effect of depth on gaze patches:

\begin{itemize}
    \item Gaze is collected within a 3D environment, therefore, the tasks should elicit interaction with objects across a wide range of distances (near and far). 
    \item The scene should be comprised of virtual and real-world objects. Object geometry should vary from simple (e.g., cube) to complex (e.g., plant) structures, to achieve a higher diversity in the gaze data set.
    \item The measured eye-to-object distance should be verifiable, for example, by placing distance markers on the scene's ground or wall.
\end{itemize}

\paragraph{Scene Description}

Figure~\ref{fig:setup} illustrates our scene setup and shows the view from the HoloLens 2. 
The scene consisted of six objects: three physical and three virtual ones. We placed the objects in alternating depths in a row on a table. 
The virtual objects comprised a potted plant (\textbf{V1}), the head of a sculpture (\textbf{V2}), and a canister (\textbf{V3}). 
The physical objects included a potted Aloe Vera plant (\textbf{R1}), a guitar placed on a stand (\textbf{R2}), and an amplifier (\textbf{R3}). 

\paragraph{Task Description}

The participants were instructed to view the six objects from three different depth levels 0.5\,m, 1.5\,m, and 3\,m, which were marked on the floor using tape. 
For each depth level, the participants received the task to view the objects from left to right (\textbf{LR}): V1 $\rightarrow$ R1 $\rightarrow$ R2 $\rightarrow$ V2 $\rightarrow$ R3 $\rightarrow$ V3, and from right to left (\textbf{RL}): V3 $\rightarrow$ R3 $\rightarrow$ V2 $\rightarrow$ R2 $\rightarrow$ R1 $\rightarrow$ V1 at their own pace.
Each participant conducted both conditions twice. We had to remove one instance due to recording issues. In total, this resulted in seven scanpaths per condition.

\subsection{Scanpath Comparison} \label{subsec:scanpath_comparison}

\begin{figure*}[t]
    \centering
    \includegraphics[width=1\textwidth]{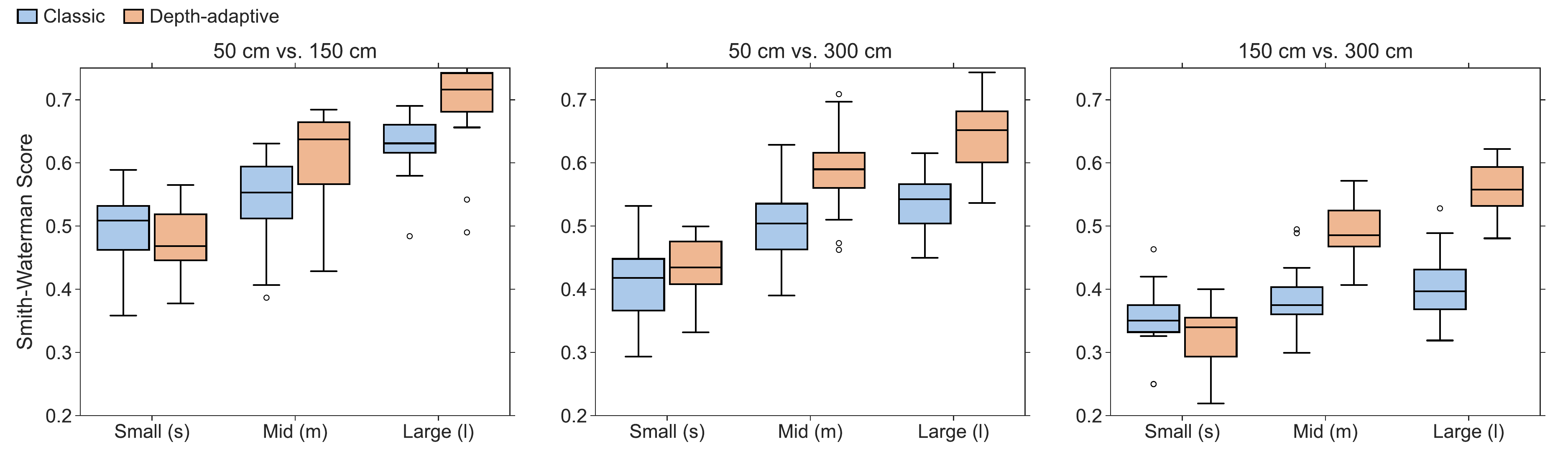}
    \caption{Smith-Waterman scores (higher are better) between scanpaths from different depth levels of 0.5\,m, 1.5\,m, and 3\,m. Similarity scores are computed between gaze patch sequences from two depth levels. Small (s), mid (m), and large (l) categories refer to different patch sizes.}
    \Description{Three boxplots (50 cm vs 150 cm; 50 cm vs 300 cm; 150 cm vs 300 cm) show the Smith-Waterman scores between the compared scanpaths. Two different scanpath representations are considered: (1) based on classic patches and (2) based on depth-adaptive patches. In each boxplot, three different patch sizes are considered: small, mid, and large. The boxplots indicate that increasing thumbnail sizes, result in a gain of scanpath similarity. Further, depth-adaptive thumbnails consistently achieve higher similarity scores on mid and large categories than classic patches.}
    \label{fig:boxplot_waterman}
\end{figure*}

Conditions \textbf{LR} and \textbf{RL} produced distinct scanpath patterns uniquely characterized by the sequence of gazed virtual and real objects.
However, recorded from different depth levels, traditional gaze thumbnails either extended beyond the focused object or clipped part of the object.
Let us now investigate the differences between our new depth-adaptive thumbnails and classic thumbnails by comparing scanpaths from the same condition across different depth levels. 
To make both methods comparable, we define three categories of thumbnail sizes, namely \textit{small~(s)}, \textit{mid~(m)}, and \textit{large~(l)}.
For the classic gaze patches, we choose the inputs $\theta_s = 2^\circ$, $\theta_m = 5^\circ$, and $\theta_l = 10^\circ$; for the  depth-adaptive gaze patches, we choose $c_s = 8$\,cm, $c_m = 20$\,cm, and $c_l = 40$\,cm.
\Cref{fig:boxplot_waterman} depicts the Smith-Waterman scores between the depth levels $0.5\,m$, $1.5\,m$, and $3.0\,m$. 
We decided to merge the similarity values from both conditions LR and RL.
There are two major observations we can extract from the results presented in \Cref{fig:boxplot_waterman}.
First, increasing the thumbnail sizes results in a gain of scanpath similarity.
Second, depth-adaptive thumbnails consistently achieve higher similarity scores on \textit{mid~(m)} and \textit{large~(l)} categories than classic patches.
We performed statistical inference between the classic thumbnails and depth-adaptive thumbnails to verify these observations.
Significant differences were noticed within the size categories of \textit{mid} and \textit{large}, but we found no significant differences in \textit{small}.
We refer to our supplemental material for details on the statistical test results. 

\subsection{Scanpath Visualization}

\begin{figure*}[h!]
    \centering
    \includegraphics[width=\textwidth]{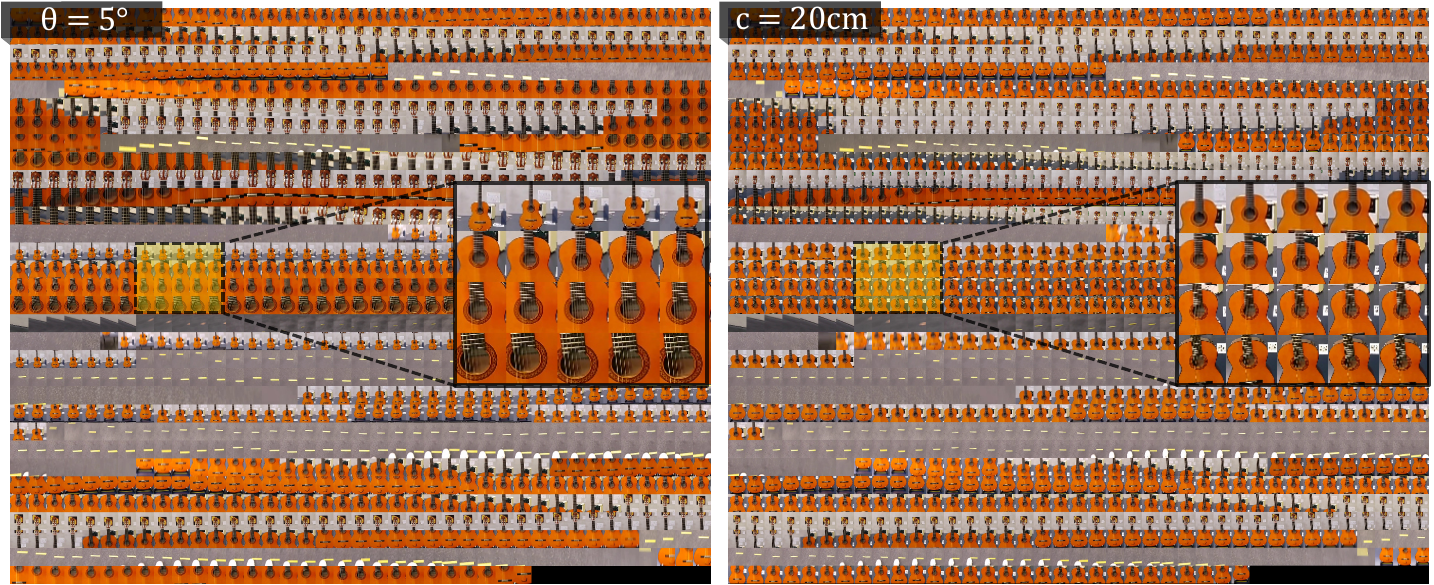}
    \caption{Gaze Stripes generated from a recording where a participant moves toward a guitar. In classic thumbnails (left), the guitar changes its apparent size as the beholder moves toward it. In depth-adaptive thumbnails, the guitar's scale remains stable across different distances (right).}
    \Description{Two Gaze Stripes of the same recording are shown where the beholder fixates on a guitar from different distances. The left Gaze Stripe is constructed from classic patches, where the apparent size of the guitar changes, causing it to clip at near distances or diminish at far distances. The right Gaze Stripe is constructed from depth-adaptive patches, where the guitar's scale remains stable across different distances.}
    \label{fig:gaze_stripes}
\end{figure*}

\begin{figure*}[h!]
    \centering
    \includegraphics[width=\textwidth]{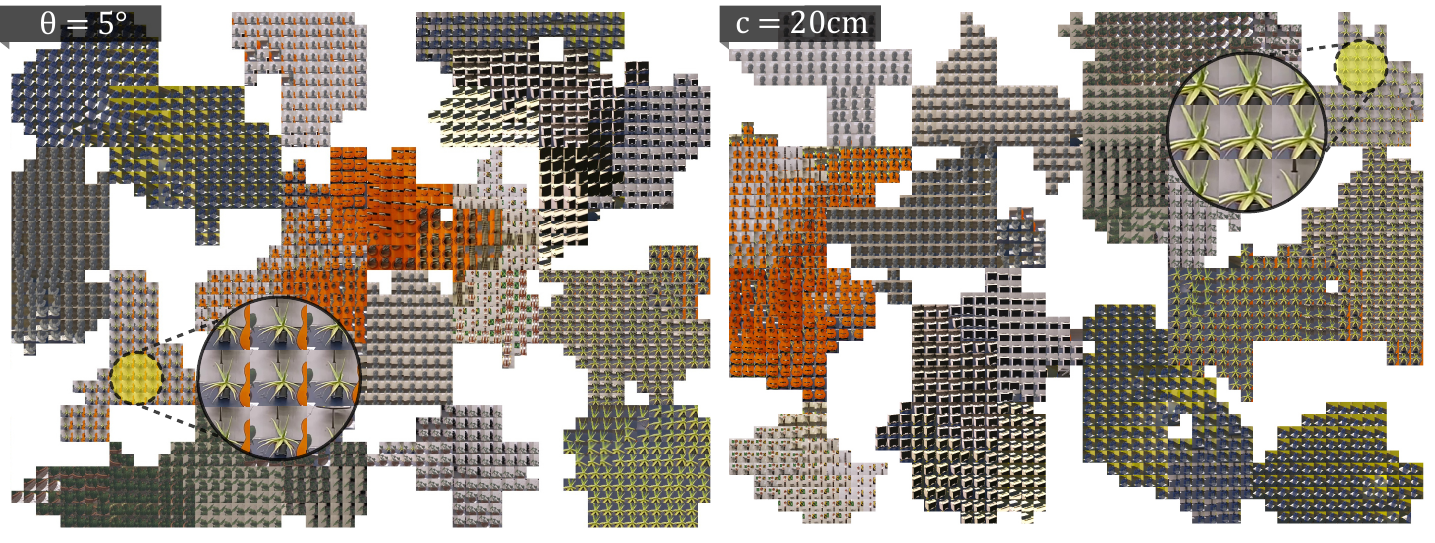}
    \caption{Gridified UMAP projection of classic thumbnails (left) and depth-adaptive thumbnails (right). In classic thumbnails, some fixations on the Aleo Vera are mapped outside the main cluster. In depth-adaptive thumbnails, all fixations on Aleo-Vera are clustered into a single region.}
    \Description{Two image-based projections of the same recording are shown where the fixates the six scene objects. The left projection is based on classic patches, where some fixations concerning the Aloe Vera plant get scattered outside its main cluster, located at the right in the projection. The right projection is based on depth-adaptive patches where all fixations concerning the Aloe Vera plant are clustered into a single region, located at the right in the projection.}
    \label{fig:dr_example}
\end{figure*}

In the following, we showcase how depth-adaptive thumbnails can facilitate the analysis of scanpaths using two established techniques: Gaze Stripes~\cite{Kurzhals2016} and image-based projections~\cite{Kurzhals2021}.

\paragraph{Gaze Stripes} 
\Cref{fig:gaze_stripes} depicts two Gaze Stripes generated from our dataset scene where a participant moved toward the desk while keeping their gaze centered on the guitar.
Comparing the Gaze Stripes from classical patches (left) with depth-adaptive patches (right) reveals noticeable differences.
In classical patches (left), the guitar changes its apparent size while the viewer moves toward the object, which causes clipping at near distances and diminishing at far distances. 
In depth-adaptive patches, the apparent size of the guitar remains relatively stable across different distances. 
Gaze Stripes are typically used to investigate sequential viewing patterns across multiple participants. Sometimes, the guitar becomes barely visible in classic patches due to the large eye-to-object distance. In contrast, depth-adaptive patches can lead to a more consistent scanpath representation of fixated objects, thus facilitating the visual analysis.

\paragraph{Image-based Projections}
\Cref{fig:dr_example} depicts two image-based projections of fixations of a single participant across all depth levels.
Similarly to \citet{Kurzhals2021}, $512$-D features are extracted from ResNet18~\cite{He2016} and mapped into 2D using UMAP~\cite{McInnes2020}.
We additionally employ gridifcation~\cite{Hilasaca2023} after the initial 2D projection to prevent thumbnails from overlapping in the scatterplot.
We can see that the amount of scene context influences the placement of image thumbnails in the 2D projection.
For example, in both scatterplots, the fixations on the Aleo Vera are mapped along the right edge.
However, in the classic thumbnails (left), some fixations on the Aloe Vera are scattered outside of this cluster, which is not the case in the depth-adaptive thumbnails (right).
This difference might become relevant in annotation tasks, where it is crucial to quickly identify the same object or area of interest in the 2D landscape of~thumbnails.

\section{Challenges and Limitations} \label{sec:challenges}

\begin{figure}
    \centering
    \includegraphics[width=0.8\linewidth]{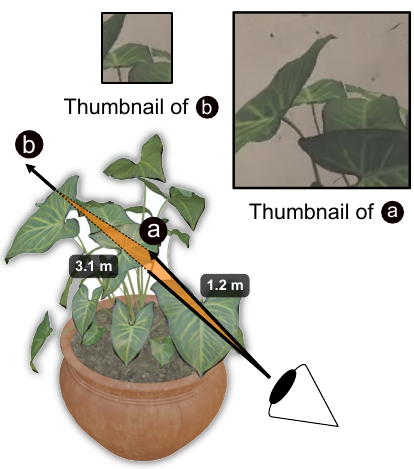}
    \caption{Unstable depth estimation can occur on virtual objects causing gaze thumbnails to change their size.}
    \Description{Example how unstable depth estimation negatively influences the depth-adaptive patches. A gaze ray passes through the potted plant (a virtual object) due to the structure of its surface mesh: the plant has wide but sparse leaves (many holes).}
    \label{fig:robustness}
\end{figure}

In the following, we discuss some of the challenges and limitations we identified in depth-adaptive thumbnails.
\paragraph{Acquiring Depth Estimation}
Modern AR headsets typically use a combination of techniques to acquire depth information, which is needed to compute the eye-to-object distance. Among such technologies are stereoscopic cameras and time-of-flight sensors. Acquiring accurate depth estimation from the current generation of mobile eye trackers (i.e., glasses) is challenging since many devices lack depth-sensing capabilities.

\paragraph{Robustness of Depth Estimation} 
There are two potential difficulties with ray-based depth estimation: (1) When a gaze ray hits the object's boundary, a small shift in gaze estimate may cause the gaze ray to miss the object mesh and intersect with far-off objects at different depths. (2) If the hit object has surface geometry that is not smooth, a small shift of the gaze ray may lead to a different eye-to-gaze distance. 
\Cref{fig:robustness} exemplifies the first issue on the plotted plant virtual object (\textbf{V1}).
In this example, the gaze ray \circled{a} hits the leaves and returns a depth estimate of $1.2$\,m, whereas \circled{b} returns $3.1$\,m since the ray passes through the object. 
This sensitivity is also reflected in the respective depth-adaptive thumbnails.

\paragraph{Image Feature Selection}
Local alignment scores and image-based projection rely on image features, which are typically extracted during preprocessing. Feature representations differ in their robustness with respect to changes in illumination, scale, or shifts. The robustness of feature representation not only depends on the particular model employed but also on the specifics of extracting features from the network. For example, extracting features from multiple layers offers more robustness than from a single layer in VGG-16~\cite{Chen2023}. Therefore, the difference between classic patches and depth-adaptive patches will likely vary depending on the employed feature representations.

\section{Conclusion} \label{sec:conclusion}
We presented a new technique for image-based analysis of gaze data from mobile eye tracking devices by adjusting the angle for the extracted thumbnails around the point of regard. 
The resulting zoom effect leads to more consistent thumbnails that improve numerous techniques including scanpath comparison, AOI classification, and data visualization.
We argue that depth-adaptive patches can be viable alternatives to classic patches, in particular, in situations with considerable relative movements between participants and the focused objects.
One possible scenario is smooth pursuits on moving objects, like a virtual bird flying within an AR Scene.
Our evaluation indicates that depth-adaptive thumbnails can be advantageous in such scenarios, which is verified by our quantitative analysis of the local alignment scores from the Smith-Waterman algorithm.
It should be noted that our evaluation is based on four participants. Thus, our results are preliminary, and a more extensive evaluation with more users and recordings is required.
We experienced that differences between the depth-adaptive thumbnails and classical thumbnails were particularly pronounced in the Gaze Stripes visualization and image-based projections.

In the future, we plan to test our approach for interactive applications with real-world objects. This will require an alternative depth estimation that is different from the one applied in this work. We will assess alternative approaches to perform live depth estimation in the environment and investigate how depth-adaptive thumbnails will affect live classification.

In summary, we think that depth-adaptive thumbnails have the potential to improve image-based gaze analysis in many ways. 
Especially for comparing scanpaths without having to annotate the data beforehand and for more efficient automatic classification of gaze on different AOIs, the presented technique could be deployed with minor changes to existing setups. 
We see the main challenge in equipping new hardware generations with appropriate sensors for depth estimation. Alternatively, vergence-based approaches and depth estimation via machine learning models might be sufficient to provide the required information for depth-adaptive thumbnails.

\begin{acks}
This work was funded by the Deutsche Forschungsgemeinschaft (DFG, German Research Foundation) -- Project-ID 449742818, Project-ID 251654672 -- TRR 161 (Project B01), and under Germany’s Excellence Strategy -- EXC 2120/1 -- 390831618.
\end{acks}

\bibliographystyle{ACM-Reference-Format}
\bibliography{sample-base}


\renewcommand{\keywordsname}{}    
\settopmatter{printccs=false}


\renewcommand{\abstractname}{}
\renewenvironment{abstract}{}{}
\renewenvironment{translatedabstract}{}{}

\title{Supplemental Material: How Deep Is Your Gaze? Leveraging Distance in Image-Based Gaze Analysis}
\maketitle
\section{Levenshtein Distances}

\begin{figure*}[t]
    \centering
    \includegraphics[width=1\textwidth]{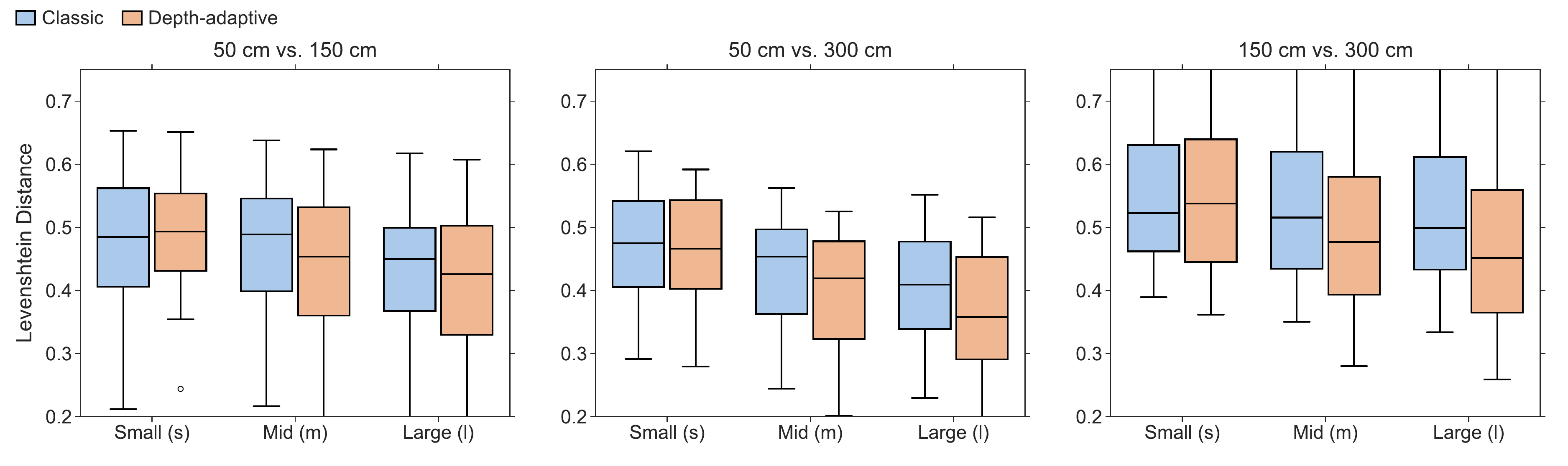}
    \caption{Levenshtein distances (lower are better) between scanpaths from different depth levels of 0.5\,m, 1.5\,m, and 3\,m. Distances are computed between gaze patch sequences from two depth levels. Small (s), mid (m), and large (l) categories refer to different patch sizes.}
    \label{fig:boxplot_levenshtein}
\end{figure*}

Similarly to the Smith-Waterman alignment, we compared classic patches with depth-adaptive patches based on the Levenshtein distances between the extracted image features. The results are presented in Figure~\ref{fig:boxplot_levenshtein}.

\section{Test Statistics on Waterman-Smith Scores}

In the following, we show the results of two-sided Student's \textit{t}-tests and two-sided Wilcoxon signed-rank tests.
The tests are performed on the Smith-Waterman alignment scores between the classic patches and depth-adaptive patches as described in our paper.
The results are organized by the thumbnail size categories \textit{Small}, \textit{Mid}, and \textit{Large}.

\subsection{Small (s)}

\paragraph{50 cm vs. 150 cm}
\begin{itemize}
\item Shapiro-Wilk test for normality (\textit{classic}): p > 0.05 (W = 0.9520, p = 0.59147)
\item Shapiro-Wilk test for normality (\textit{depth-adaptive}): p > 0.05 (W = 0.9555, p = 0.64824)
\item t-test (\textit{classic}, \textit{depth-adaptive}): p > 0.05 (F = 0.5580, p = 0.58161)
\item Wilcoxon signed-rank test (\textit{classic}, \textit{depth-adaptive}): p > 0.05 (F = 38.0000, p = 0.39099)
\end{itemize}
\paragraph{50 cm vs. 300 cm}
\begin{itemize}
\item Shapiro-Wilk test for normality (\textit{classic}): p > 0.05 (W = 0.9711, p = 0.89118)
\item Shapiro-Wilk test for normality (\textit{depth-adaptive}): p > 0.05 (W = 0.9197, p = 0.21782)
\item t-test (\textit{classic}, \textit{depth-adaptive}): p > 0.05 (F = -1.0015, p = 0.32580)
\item Wilcoxon signed-rank test (\textit{classic}, \textit{depth-adaptive}): p > 0.05 (F = 24.0000, p = 0.07849)
\end{itemize}
\paragraph{150 cm vs. 300 cm}
\begin{itemize}
\item Shapiro-Wilk test for normality (\textit{classic}): p > 0.05 (W = 0.9337, p = 0.34315)
\item Shapiro-Wilk test for normality (\textit{depth-adaptive}): p > 0.05 (W = 0.9516, p = 0.58546)
\item t-test (\textit{classic}, \textit{depth-adaptive}): p > 0.05 (F = 1.1129, p = 0.27594)
\item Wilcoxon signed-rank test (\textit{classic}, \textit{depth-adaptive}): p > 0.05 (F = 36.0000, p = 0.32581)
\end{itemize}

\subsection{Mid (m)}

\paragraph{50 cm vs. 150 cm}
\begin{itemize}
\item Shapiro-Wilk test for normality (\textit{classic}): p > 0.05 (W = 0.8901, p = 0.08112)
\item Shapiro-Wilk test for normality (\textit{depth-adaptive}): p < 0.05 (W = 0.8047, p = 0.00571)
\item t-test (\textit{classic}, \textit{depth-adaptive}): p < 0.05 (F = -2.1617, p = 0.04002)
\item Wilcoxon signed-rank test (\textit{classic}, \textit{depth-adaptive}): p < 0.05 (F = 0.0000, p = 0.00012)
\end{itemize}
\paragraph{50 cm vs. 300 cm}
\begin{itemize}
\item Shapiro-Wilk test for normality (\textit{classic}): p > 0.05 (W = 0.9902, p = 0.99962)
\item Shapiro-Wilk test for normality (\textit{depth-adaptive}): p > 0.05 (W = 0.9471, p = 0.51720)
\item t-test (\textit{classic}, \textit{depth-adaptive}): p < 0.05 (F = -3.1903, p = 0.00369)
\item Wilcoxon signed-rank test (\textit{classic}, \textit{depth-adaptive}): p < 0.05 (F = 0.0000, p = 0.00012)
\end{itemize}
\paragraph{150 cm vs. 300 cm}
\begin{itemize}
\item Shapiro-Wilk test for normality (\textit{classic}): p > 0.05 (W = 0.9122, p = 0.16966)
\item Shapiro-Wilk test for normality (\textit{depth-adaptive}): p > 0.05 (W = 0.9682, p = 0.85193)
\item t-test (\textit{classic}, \textit{depth-adaptive}): p < 0.05 (F = -5.5544, p = 0.00001)
\item Wilcoxon signed-rank test (\textit{classic}, \textit{depth-adaptive}): p < 0.05 (F = 3.0000, p = 0.00061)
\end{itemize}

\subsection{Large (l)}

\paragraph{50 cm vs. 150 cm}
\begin{itemize}
\item Shapiro-Wilk test for normality (\textit{classic}): p < 0.05 (W = 0.8522, p = 0.02383)
\item Shapiro-Wilk test for normality (\textit{depth-adaptive}): p < 0.05 (W = 0.7811, p = 0.00294)
\item t-test (\textit{classic}, \textit{depth-adaptive}): p < 0.05 (F = -2.4240, p = 0.02261)
\item Wilcoxon signed-rank test (\textit{classic}, \textit{depth-adaptive}): p < 0.05 (F = 2.0000, p = 0.00037)
\end{itemize}
\paragraph{50 cm vs. 300 cm}
\begin{itemize}
\item Shapiro-Wilk test for normality (\textit{classic}): p > 0.05 (W = 0.9564, p = 0.66393)
\item Shapiro-Wilk test for normality (\textit{depth-adaptive}): p > 0.05 (W = 0.9577, p = 0.68521)
\item t-test (\textit{classic}, \textit{depth-adaptive}): p < 0.05 (F = -5.4715, p = 0.00001)
\item Wilcoxon signed-rank test (\textit{classic}, \textit{depth-adaptive}): p < 0.05 (F = 0.0000, p = 0.00012)
\end{itemize}
\paragraph{150 cm vs. 300 cm}
\begin{itemize}
\item Shapiro-Wilk test for normality (\textit{classic}): p > 0.05 (W = 0.9578, p = 0.68719)
\item Shapiro-Wilk test for normality (\textit{depth-adaptive}): p > 0.05 (W = 0.9225, p = 0.23918)
\item t-test (\textit{classic}, \textit{depth-adaptive}): p < 0.05 (F = -7.5592, p = 0.00000)
\item Wilcoxon signed-rank test (\textit{classic}, \textit{depth-adaptive}): p < 0.05 (F = 0.0000, p = 0.00012)
\end{itemize}

\end{document}